\documentclass[twocolumn,showpacs,pra,aps]{revtex4}
\usepackage{array}
\usepackage{amsmath}
\usepackage{graphicx}
\usepackage{amstext}
\usepackage{psfig}
\usepackage{amsfonts}
\begin{document}

\title{Using weak nonlinearity under decoherence for
macroscopic-entanglement generation and quantum computation}
  
\author{Hyunseok Jeong}


\affiliation{Centre for Quantum Computer Technology, 
Department of Physics, University of Queensland, St
Lucia, Qld 4072, Australia}

\date{\today}

\begin{abstract}
Recently, there have been several suggestions 
that weak Kerr nonlinearity can be used for generation of macroscopic 
superpositions and entanglement and for linear optics quantum computation.
However, it is not immediately clear that this approach can 
overcome decoherence effects.
Our numerical study in this paper shows that 
nonlinearity of weak strength could be useful
for macroscopic entanglement generation and quantum gate operations
in the presence of decoherence.
We suggest specific values for real experiments based on our analysis.
Our discussion shows that the generation of macroscopic entanglement
using this approach is within reach of current technology.
\end{abstract}


\maketitle

Strong nonlinear effects, if available, could be very useful for the
generation of macroscopic superpositions and 
entanglement and for quantum information processing. For example,
it is well known that macroscopic superpositions
and entanglement could be generated in strong Kerr nonlinear media
\cite{Yurke}.
An optical quantum computer could be realized if strong 
nonlinearity was available \cite{NC}.
However, it is extremely hard to obtain such strong nonlinear effects
using nonlinear media.
Nonlinear effects
in  existing media are extremely weak compared
with the required level for generation of
macroscopic superpositions and entanglement or for
quantum information processing.

Recently, it was suggested that weak Kerr nonlinearity can
still be used for production of macroscopic 
superpositions and entanglement \cite{Enk03,Jeong04}
and for linear optics quantum computation \cite{NM04}. 
Nemoto and Munro showed that weak nonlinearity could be used
to generate entangled states and used for linear optics quantum computation \cite{NM04,addedref},
which could also be used for the generation of macroscopic superposition \cite{Gerry,MunroP}. 
A key element of their scheme is a cross-Kerr interaction between
a coherent state and a single-photon qubit. The phase of the initial coherent state 
changes by a certain amount when the qubit is $|1\rangle$, while it remains the same
when the qubit is $|0\rangle$. An important point of this scheme is that the amplitude
of the initial coherent state should become arbitrarily large to make
the required nonlinear strength arbitrarily weak. 
The initial coherent state can easily gain an adequate
amount of phase shift by $|1\rangle$ when $\alpha$ becomes
very large.
This effect could be used to generate quantum entanglement
with a very weak nonlinearity.
The other schemes 
based on the singe-mode Kerr effect
by van Enk \cite{Enk03} and Jeong {\it et al.} \cite{Jeong04}
use implicitly
the same principle of using a coherent state of a
large amplitude:
all these schemes 
\cite{Enk03,Jeong04,NM04,addedref}
use initial coherent states of large amplitudes 
in order to make the required separation between the component coherent
states in the phase space with weak nonlinearity.

However, decoherence effects 
{\it during} the entanglement generation process
in nonlinear media have not been investigated
in these references \cite{Enk03,Jeong04,NM04} 
in spite of the uncertainty 
of validity of such techniques
under a real dissipative environment.
As it was mentioned,  
the amplitude of the initial coherent state must increase at
the cost of decreasing the required nonlinear strength (or 
required interaction time in 
a nonlinear medium).
The requirement of a large initial amplitude
might reduce the coherence time of the evolving state,
i.e., an entangled system 
with a large initial amplitude might lose its quantum coherence more rapidly
than an entangled system with a small initial amplitude did.
It is thus unclear whether the decoherence effects (particularly dephasing) 
could be overcome by this approach
when generating entanglement and operating quantum gates.
In this paper,
to answer this question,
we investigate the effect of decoherence 
when the idea of using weak nonlinearity
\cite{NM04}
is applied to Gerry's scheme \cite{Gerry}
to generate a macroscopic superposition (so-called Schr\"odinger cat state) 
 in a dissipative environment. 

The interaction Hamiltonian of cross-Kerr nonlinearity 
between modes 1 and 2 is
$H_{K}=\hbar \chi a_1^\dagger a_1 a_2^\dagger a_2$,
where $a$ ($a^\dagger$) represents the annihilation (creation) operator.
The interaction between a coherent state, $|\alpha\rangle_2$, and
a single-photon qubit, e.g.,
$|\psi\rangle_1=(|0\rangle_1+|1\rangle_1)/\sqrt{2}$, is described as
\begin{eqnarray}
U_K(t)|\psi\rangle_1|\alpha\rangle_2&=&e^{i H_{K}t/\hbar}\frac{1}{\sqrt{2}}
(|0\rangle_1+|1\rangle_1)|\alpha\rangle_2\nonumber\\
&=&\frac{1}{\sqrt{2}}(|0\rangle_1|\alpha\rangle_2+|1\rangle_1|\alpha e^{i\theta}\rangle_2),
\end{eqnarray} 
where
$|0\rangle$ ($|1\rangle$) is the vacuum (single-photon) state,
$\alpha$ is the amplitude of the coherent state,
and $\theta=\chi t$ with the interaction time $t$.
If $\theta$ is $\pi$ and one measures out mode 1 on a superposed basis 
$(|0\rangle\pm|1\rangle)/\sqrt{2}$, a macroscopic superposition state,
$|\Phi_\pm\rangle=(|\alpha\rangle\pm|-\alpha\rangle)/\sqrt{M_\pm}$ is created,
where $M_\pm=2\pm 2\exp[-2|\alpha|^2]$. 
A macroscopic entanglement can be simply
generated at a beam splitter with such a state.

\begin{figure}
\centerline{\scalebox{0.55}{\includegraphics{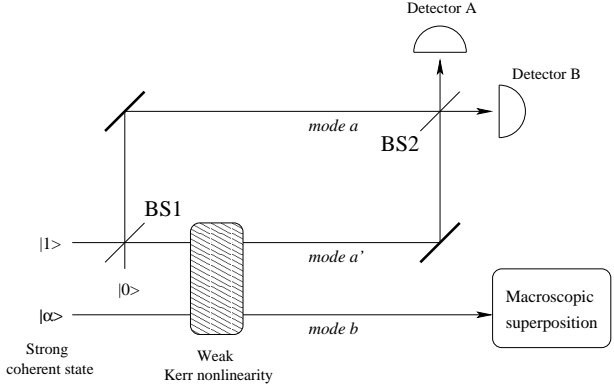}}}
 \caption{A schematic of Gerry's scheme \cite{Gerry} combined with 
the idea of using weak nonlinearity \cite{NM04}. 
   } \label{fig:gerry}
\end{figure}

Using the dual rail logic,
where the logical qubit basis is defined as $|0_L\rangle\equiv|1\rangle\otimes|0\rangle$
and $|1_L\rangle\equiv|0\rangle\otimes|1\rangle$, the above process can be 
efficiently realized.
For example, as shown in Fig.~\ref{fig:gerry},
Gerry's scheme \cite{Gerry} can be directly linked 
to the idea of using weak nonlinearity \cite{NM04} so that weak
cross-Kerr nonlinearity can be used with 
a single photon, a coherent state,
two photodetectors and two beam splitters to generate
a macroscopic superposition.
If detector A (detector B) clicks, 
a macroscopic superposition state $|\Phi_-\rangle$ ($|\Phi_+\rangle$)
 is obtained
at mode $b$ in Fig.~\ref{fig:gerry}.
Remarkably, it 
is clear from Fig.~\ref{fig:gerry} that this approach is robust against 
inefficiency of the single photon source, loss of the single photon
and inefficiency of the photodetectors.
Those factors will cause the photodetectors to be
silent and such cases can simply be discarded.
Therefore, these will only make the deterministic property of the scheme
non-deterministic
but will not affect the quality of the obtained macroscopic superposition state.

The main problem here is that a very large nonlinear effect,
i.e. a very large $\theta$,
is required to gain a large separation between two coherent component states.
It was pointed out that an optical fiber of about $3,000km$
is required for $\theta=\pi$
for an optical frequency of $\omega\approx5\times10^{14}$ rad/sec
using currently available Kerr nonlinearity \cite{SandersMilburn}.
In such a case,
 the state after the nonlinear
interaction will be completely decohered
because of the significant losses in the fiber. 
In order to circumvent this problem, a large
initial amplitude $\alpha$ can be used with a short interaction time $t$.  
If $\alpha$ is very large, the same amount of separation can be obtained in the phase
space even though $\theta$ ($=\chi t$) is much smaller than $\pi$.

Now we consider 
the decoherence effects in the Kerr medium.
The decoherence effects
can be induced by solving the master equation \cite{Phoenix}
\begin{equation}
\label{master-eq}
  {\partial \rho \over \partial t}=\hat{J}\rho +\hat{L}\rho~;~\hat{J}\rho=\gamma a\rho a^\dag,~~
  \hat{L}\rho=-{\gamma \over 2}(a^\dag a\rho +\rho a^\dag a)
\end{equation}
where $\gamma$ is the energy decay rate. The formal solution of the
master equation (\ref{master-eq}) can be written as
$\rho(t)=\exp[(\hat{J}+\hat{L})t]\rho(0)$,
which leads to the solution for the initial element
$|\alpha\rangle\langle\beta|$
\begin{eqnarray}
  \exp[(\hat{J}+\hat{L})t]|\alpha\rangle\langle\beta|\equiv{\cal\tilde D}(t)|\alpha\rangle\langle\beta|
~~~~~~~~~~~~~~~~~~~~~~~~\nonumber\\
  =
\exp[-\frac{1}{2}(1-e^{-\gamma t})(|\alpha|^2+|\beta|^2)+\alpha\beta^*] 
  |A \alpha \rangle\langle A\beta |,
\label{solution-master}
\end{eqnarray}
where $A=e^{-\gamma t/2}$.

We numerically assess the decoherence effects as follows.
As we have explained, a photon loss at mode 1 (modes
$a$ and $a^\prime$ in Fig.~\ref{fig:gerry}) will
simply cause the success probability
to be less than 1.
However, energy loss of the coherent state part (mode $b$ in Fig.~\ref{fig:gerry})
in the nonlinear medium 
should be seriously considered since it 
will cause decoherence (dephasing) of the obtained state.
Note that the average energy loss per time increases as the initial energy gets larger. 
This will cause a more rapid destruction of quantum coherence
for a large $\alpha$.
The decoherence process ($\cal\tilde D$) will occur simultaneously 
with the unitary evolution by the Kerr effect ($\cal\tilde U$) of
the input state by the interaction Hamiltonian $H_K$. 
This process may be modeled as follows.
One may assume that
$\cal\tilde U$ occurs for a short time $\Delta t$, and then $\cal\tilde D$ occurs for another $\Delta t$.
In other words, $\cal\tilde U$ and $\cal\tilde D$
continuously take turn for a short time in the nonlinear medium.
By taking $\Delta t$ arbitrarily small, one can obtain a good approximation
of this process for a certain time $t$ ($=N\Delta t$), where $N$ is an integer number.
Let us set 
$\Delta\theta=\chi\Delta t=\pi/N$,
then a larger $N$ will result in a better approximation.

We now use our model to analyze the behavior of the
coherent state interacting with a logical qubit in the  Kerr medium
shown in Fig.~\ref{fig:gerry}.
The first beam splitter, BS1, and the single photon prepares
the logical qubit state, $(|0_L\rangle+|1_L\rangle)/\sqrt{2}
\equiv(|1\rangle_a|0\rangle_{a^\prime}+|0\rangle_a|1\rangle_{a^\prime})/\sqrt{2}$. 
The total initial state after BS1 but
before the Kerr interaction in Fig.~\ref{fig:gerry} is
\begin{equation}
\rho(t=0)
=
\frac{1}{2}\big(|0_L\rangle\langle 0_L|+
|0_L\rangle\langle 1_L|\nonumber
+
|1_L\rangle\langle 0_L|+
|1_L\rangle\langle 1_L|\big)\otimes|\alpha\rangle\langle\alpha|.
\end{equation}
Let us fist consider the evolution of
the second term, 
$|0_L\rangle\langle 1_L|\otimes|\alpha\rangle\langle\alpha|$.
After time $t$ $(=N\Delta t)$ in the nonlinear medium, it evolves to 
\begin{equation}
\Big\{{\cal\tilde D}(\Delta t){\cal\tilde U}(\Delta t)\Big\}^N
|0_L\rangle\langle 1_L|\otimes|\alpha\rangle\langle\alpha|
=C|0_L\rangle\langle 1_L|\otimes
|A\alpha\rangle\langle A\alpha e^{i\theta}|
\end{equation}
where ${\cal\tilde U}(\Delta t)\rho\equiv U_K(\Delta t)\rho U_K^\dagger(\Delta t)$
and
\begin{eqnarray}
C=\exp[-\alpha^2(1-e^{-\gamma(t/N)})~~~~~~~~~~~~~~~~~~~~~~~~~~~~\nonumber\\
\sum_{n=1}^N\exp[-\gamma(t/N)]^{(n-1)}(1-\exp[I\chi n(t/N)])]
\end{eqnarray}
where 
$\Gamma=\chi/\gamma$ 
and  $\alpha$ is assumed to be real without losing generality.
Here we have defined $C$ as ``coherence parameter'' since it determines
 the degree of decoherence (dephasing) for the resulting macroscopic superposition state.
The amplitude parameter $A$ has been defined to quantify the average
energy loss.
The second beam splitter, BS2, and detectors A and B in Fig.~\ref{fig:gerry} perform
a measurement onto the superposed basis states $(|0_L\rangle\pm|1_L\rangle)/\sqrt{2}$.  
The macroscopic superposition state obtained 
by the measurement is
\begin{eqnarray}
\rho_\pm(t)={\cal N}_\pm(|A\alpha\rangle \langle A\alpha|\pm
C|A\alpha\rangle\langle A\alpha e^{i\theta}|
~~~~~~~\nonumber\\ 
\pm
 C^*|A\alpha e^{i\theta}\rangle\langle A\alpha|+|A\alpha 
e^{i\theta}\rangle\langle A\alpha e^{i\theta}|)
\label{eqn-a}
\end{eqnarray}
where ${\cal N}_\pm$ are the normalization factors.
State $\rho_+(t)$ ($\rho_-(t)$) is obtained when 
detector B (detector A) clicks in Fig.~\ref{fig:gerry}.
If $|C|=1$ (and $A\neq 0$) 
the state $\rho_\pm(t)$ is a pure superposition of coherent states
while if $|C|=0$ it is simply a statistical mixture of two coherent states.
 Let us assume that
one wishes to gain $\theta=\pi$ 
with the amplitude of the initial coherent state is $\alpha_0$
so that a macroscopic superposition state $|\alpha_0\rangle\pm|-\alpha_0\rangle$
(unnormalized) could be obtained (without decoherence). In this case, an interaction time
$t=\pi/\chi$ is required. 
However, if the amplitude of the initial coherent state is  
$\alpha$ ($>\alpha_0$), the required interaction time $t$ is
obtained from the equation,
$|\alpha e^{i\chi t}-\alpha|=2\alpha_0$,
which can be derived from a simple geometric analysis. 
The required interaction time is then 
$t(\alpha,\alpha_0,\chi)=\cos^{-1} [1-2\alpha_0^2/\alpha^2]/\chi$.

Note that the state (\ref{eqn-a}) can be changed into a 
symmetric form in the phase space, i.e., $A\alpha\rightarrow\delta$
and $A\alpha e^{i\theta}\rightarrow-\delta$.
This can be done by applying the displacement operator,
$D(x)=\exp(xa^\dag+x^* a)$,
where $a$ and $a^\dag$ are annihilation and creation operators.
The displacement operation can be performed using a strong coherent field (an additional local oscillator)
and a biased beam splitter. This may be required for quantum information processing 
but it does not make any essential difference in our discussions.

\begin{figure}
\centerline{\scalebox{0.6}{\includegraphics{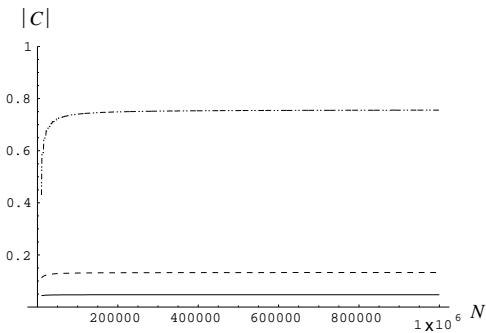}}}
\caption{This figure shows that our approach gives a very good
 approximation throughout the whole range of $\alpha$ that we consider in this paper.
The coherence parameter $|C|$ for a separation $|\alpha e^{i\chi t}-\alpha|=2\alpha_0=6$ has been plotted
for $\alpha=300$ (solid line), $\alpha=1000$ (dashed line) and $\alpha=10000$ (double-dot-dashed line).
   } \label{fig:approx}
\end{figure}

\begin{figure}
\centerline{\scalebox{0.6}{\includegraphics{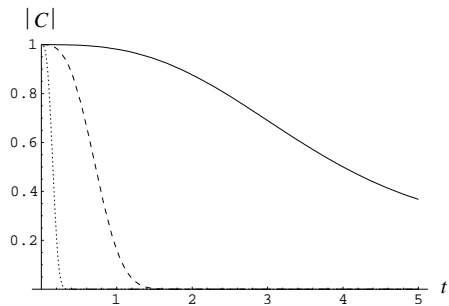}}}
 \caption{
The decrease of coherence parameter $|C|$ against time $t$. 
 $\alpha=3$ (solid line), $\alpha=30$ (dashed line),
  $\alpha=300$ (dotted line). 
Decoherence occurs faster as the initial amplitude gets larger.
$\chi/\gamma=0.0125$.
   } \label{fig:time}
\end{figure}

\begin{figure}
\centerline{\scalebox{0.6}{\includegraphics{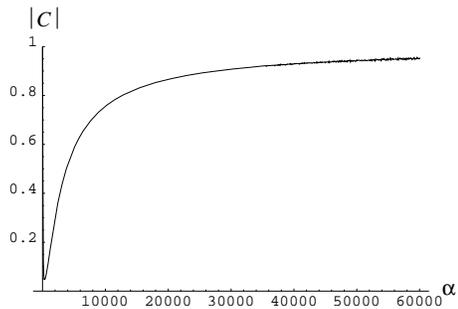}}}
 \caption{
The coherence parameter $|C|$
against the initial amplitude $\alpha$ 
for the same separation ($\alpha_0=3$).
The decoherence effect diminishes as $\alpha$ gets larger. 
See text for details.
 } \label{fig:coherence}
\end{figure}

In our numerical calculation, we have chosen $N=10^6$, i.e.,
$\Delta \theta=\pi/10^6$.
It is clear from Fig.~\ref{fig:approx} that this value gives a very good approximation
for the whole range of $\alpha$ in our study because
$|C|$ rapidly converges as $N$ increases.
Our numerical results can be summarized as follows.
The first effect is that the same amount of separation between the coherent states
is gained in a shorter time for a larger initial amplitude, which we already learned
from Nemoto and Munro \cite{NM04}.
The second effect is that decoherence occurs faster as the initial amplitude gets larger as shown in Fig.~\ref{fig:time}, 
which could be expected intuitively.
Our observation is that
the first effect overcomes the second one so that decoherence effects diminish
as the initial amplitude gets larger
for the same amount of separation:
Fig.~\ref{fig:coherence} shows
that the coherence parameter $|C|$ increases as the initial amplitude $\alpha$ increases for a same separation.
In what follows we present 
the detailed values obtained from our numerical study.
One may need an optical fiber of about $3,000km$
for $\theta=\pi$
using currently available Kerr nonlinearity
\cite{SandersMilburn}.
We choose $\Gamma=0.0125$ 
that the amplitude will reduce as $A\approx0.533$ for $15km$ while 
$\theta=\pi$ is obtained for $3,000km$.
This corresponds ~0.364dB/km of signal loss, 
which is a typical value for commercial fibers used
for telecommunication and easily achieved using current technology
\cite{fiberloss1,fiberloss2}.
Note that signal losses in some pure silica core fibers are even less than 0.15dB/km 
\cite{fiberloss2}.
If the required amplitude for the obtained cat state is $\alpha_0=3$ and
the initial amplitude is also $\alpha=3$ (so that $\theta=\pi$),
the amplitude parameter is $A\approx2.7\times10^{-55}$, i.e.,
the resulting state will be virtually the vacuum 
\footnote{In this case ($\alpha=\alpha_0=3$) the nonlinearity $\approx 10^4$
times larger than the currently existing value is required for $A>0.9$ and $|C|\approx 0.8$.}.
If $\alpha=300$, the amplitude parameter is $A=0.45$ so that the ``effective"
amplitude, $|A\alpha-A\alpha^{i\theta}|/2$, of the cat state
calculated from the separation between the two coherent states
is $\approx1.35$. In this case the coherence parameter is
$|C|\approx0.047$, i.e., the state is almost completely decohered.
If $\alpha=3000$, the effective amplitude
is $\approx2.76$ and the coherence parameter is
$|C|\approx0.43$. 
 If $\alpha=30000$, the effective amplitude
is $\approx2.97$ and the coherence parameter is
$|C|\approx0.91$, i.e., the resulting state will be close to a pure
macroscopic superposition state.
Therefore, in order to obtain $|C|>0.9$ for 
the (effective) amplitude $|\alpha|\approx 3$ of the cat state,
one needs the initial coherent state of $\alpha=30000$.
In this case, an optical fiber of only about $190m$ will be required.

One can simply produce macroscopic entanglement using an additional
50:50 beam splitter 
on the state produced in Fig.~\ref{fig:gerry} \cite{Sanders}.
The state (\ref{eqn-a}) after 
this additional beam splitter becomes
\begin{eqnarray}
\rho^E_\pm(t)={\cal N}_\pm(|
\beta,\beta\rangle \langle \beta,\beta|\pm
C|\beta,\beta\rangle\langle \beta^\prime,\beta^\prime|~~~~\nonumber\\
\pm C^*|\beta^\prime,\beta^\prime\rangle\langle \beta,\beta|+
|\beta^\prime,\beta^\prime\rangle\langle\beta^\prime,\beta^\prime|)
\label{eqn-bb}
\end{eqnarray}
where $\beta=A\alpha/\sqrt{2}$, $\beta^\prime=A\alpha e^{i\theta}/\sqrt{2}$
and $|\beta,\beta\rangle=
|\beta\rangle|\beta\rangle$.
Applying displacement operators $D_1(x)D_2(x)$, 
this state can be transformed to a symmetric form as 
\begin{eqnarray}
\rho^{E^\prime}_\pm(t)={\cal N}_\pm(|
\delta,\delta\rangle \langle \delta,\delta|\pm
C^\prime|\delta,\delta\rangle\langle -\delta,-\delta|
~~~~~~~~~~\nonumber\\
\pm {C^\prime}^*|-\delta,-\delta\rangle\langle \delta,\delta|+
|-\delta,-\delta\rangle\langle -\delta,-\delta|)
\label{eqn-cc}
\end{eqnarray}
where 
$x=-(\beta+\beta^\prime)/2$, 
$\delta=(\beta-\beta^\prime)/2$ and 
$C^\prime=C\exp[2A^2\alpha^2 i \sin \theta]$.
Note that the state (\ref{eqn-bb}) and the state (\ref{eqn-cc}) contain the same amount
of entanglement because local unitary transformations do not increase nor decrease the
amount of entanglement.  
The state (\ref{eqn-cc})
can be represented in a $2\times 2$ Hilbert space
by defining the basis as $\{|\Phi_+\rangle,|\Phi_-\rangle\}$.
The state $\rho^{E^\prime}_+$ in Eq.~(\ref{eqn-cc}) in the 
$2\times 2$ Hilbert space spanned by the new basis
is then
\begin{equation}
\rho^{E^\prime}_+  =
\frac{1}{K+2R+Z}\left(
  \begin{array}{cccc}
  K & V & V & D \\
  -V & R & R & W\\
  -V & R & R & W\\
  D & -W & -W & Z
\end{array}\right) \label{eq:matrix},
\end{equation}
where
$K=M_+^2(2+C^\prime+{C^\prime}^*)$, 
$V=-M_+\sqrt{M_+M_-}(C^\prime-{C^\prime}^*)$,
$D=M_+M_-(2+C^\prime+{C^\prime}^*)$, 
$R=M_+M_-(2-C^\prime-{C^\prime}^*)$, 
$W=M_-\sqrt{M_+M_-}(C^\prime-{C^\prime}^*)$, 
$Z=M_-^2(2+C^\prime+{C^\prime}^*)$.
We have numerically calculated
the negative eigenvalue $\lambda_-$ of the partial transpose 
for $\rho^{E^\prime}_+$ as a measure of entanglement \cite{neg}.
In Fig.~5, one can clearly see that the degree of entanglement $E$ $(=-2\lambda_-)$ increases
as the initial amplitude becomes larger.

Our discussion clarifies that an inefficient single photon source, two inefficient detectors,
and weak nonlinearity, beam splitters and a coherent state source are required resources for
generation of macroscopic entanglement, and such a method can overcome the decoherence effect.
In the same manner, weak nonlinearity can also be used for quantum gate operations \cite{NM04} 
in the presence of decoherence.
However,
it should be noted that some effects in the nonlinear media such as phase noise
may not be negligible in real experiments.
It is also a separate problem to investigate the decoherence effect 
{\it in nonlinear media} for the other schemes \cite{Enk03,Jeong04}
with weak nonlinearity. 

{\it Note added} --- After we submitted our paper,
an update \cite{update} to Ref.~\cite{NM04} was published,
where the authors discussed decoherence of
qubits for their computation scheme.

This work was supported by the Australian
Research Council. The author acknowledges
helpful discussions with T.C. Ralph and M.S. Kim.

\begin{figure}
\centerline{\scalebox{0.6}{\includegraphics{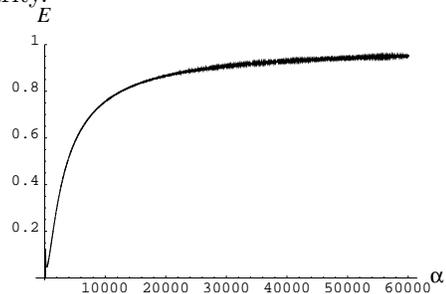}}}
 \caption{The measure of entanglement $E$ of the 
 state obtained by the
 process described in Fig.~1 against the initial amplitude $\alpha$. 
 } \label{fig:moe}
\end{figure}

\end{document}